\documentclass{optica-article}

\journal{opticajournal} % for journals or Optica Open

\articletype{Research Article}

\usepackage{lineno}
\usepackage{siunitx}
\usepackage{float}
\usepackage{caption}
\usepackage{subcaption}
\usepackage{comment}
\usepackage{fancyhdr}
\usepackage{lastpage}
\usepackage{soul,indentfirst}
\usepackage{bm}
\begin{document}

\title{Compact Optical-Resolution Photoacoustic Mircoscopy System with Reflective Objective-Based Transducer Integration}

\author{Albano Tabacchi\authormark{1}{\textsuperscript{\textdagger}}, Bhanu Pratap Singh \authormark{1}{\textsuperscript{\textdagger}{*}}, Michael Jaeger\authormark{1}, Damien Guignet \authormark{1}, Mirjam Schenk \authormark{2}, Pavel Subochev \authormark{3}, Martin Frenz \authormark{1}, and André Stefanov\authormark{1}}

\address{\authormark{1}Institute of Applied Physics, University of Bern, Sidlerstrasse 5, 3012 Bern, Switzerland\\
\authormark{2} Institute of Tissue Medicine and Pathology, University of Bern, Murtenstrasse 31, 3008 Bern, Switzerland \\
\authormark{3} Institute of Applied Physics, Russian Academy of Sciences, Nizhny Novgorod, Russia}

\noindent {\textsuperscript{\textdagger}} These authors contributed equally to this work.

\email{\authormark{*}bhanu.singh@unibe.ch} 

\section*{Abstract}

We present an optical-resolution photoacoustic microscopy (OR-PAM) system designed to overcome key limitations in conventional transducer integration within a compact microscopy configuration, while preserving high optical performance and improving acoustic detection efficiency. The system uses a reflective objective that reduces spatial constraints within the optical pathway, enabling the integration of a large-area PVDF transducer within the optical obscuration zone. The system performance was characterized through spatial resolution analysis, laser pulse energy measurement at the sample plane, and evaluation of photoacoustic signal dependence on laser pulse energy. For biological validation, OR-PAM imaging of sections from B16F10 tumors implanted in mice was performed and compared with optical microscopy and H\&E stained histological sections. The results demonstrate strong spatial correlation between photoacoustic signal intensity and melanin rich regions, confirming label-free sensitivity to endogenous optical absorbers at $532 nm$. This work establishes a compact OR-PAM imaging setup with improved optical-acoustic integration for high resolution biomedical imaging applications, with potential for future extension to multi-wavelength laser excitation.

%%%%%%%%%%%%%%%%%%%%%%%%%%  body  %%%%%%%%%%%%%%%%%%%%%%%%%%

\section{Introduction}

Photoacoustic Imaging (PAI), also referred to as Optoacoustic Imaging, is a hybrid imaging modality that combines optical excitation with ultrasound detection based on the photoacoustic effect \cite{ATTIA2019100144}\cite{mi15081007}. PAI relies on the thermoelastic effect where absorption of optical energy causes localized temperature changes, inducing thermoelastic expansion and generation of ultrasonic waves. These ultrasound waves are detected and used for image reconstruction \cite{park2024clinical}\cite{BOSSY201622}. During the past two decades, PAI has gained prominence in biomedical research, with applications in preclinical and clinical studies \cite{ATTIA2019100144}\cite{park2024clinical}\cite{zhu2024high}\cite{knieling2025primer}.

Photoacoustic Imaging comprises two principal implementations: Photoacoustic Tomography (PAT) and Photoacoustic Microscopy (PAM), with the latter being the focus of this research. PAM was first conceptualized in the 1970s \cite{knieling2025primer} and enables the visualization of intrinsic and extrinsic optical absorbers through the detection of acoustically generated signals \cite{doi:10.1126/sciadv.ado0518}. A major advantage of PAM over optical microscopy is its ability to provide label-free distinction of biological substances based on their intrinsic wavelength-dependent optical absorption, thereby enabling the imaging of endogenous chromophores such as melanin, lipids, hemoglobin, and nucleic acids \cite{doi:10.1126/sciadv.ado0518}. PAM can be further categorized into optical-resolution (OR-PAM) and acoustic-resolution (AR-PAM). OR-PAM uses focused optical excitation to achieve submicrometer lateral resolution, making it particularly suitable for high-resolution imaging of superficial tissues at depths typically below $100 \mu m$ \cite{zhu2024high}\cite{doi:10.1126/sciadv.ado0518}.

A major challenge in the design of OR-PAM systems is the integration of the ultrasonic transducer within the optical setup to enable efficient acoustic signal detection while maintaining precise optical-acoustic confocal alignment and optimizing overall imaging performance. Various approaches have been explored in the literature, each associated with distinct advantages and trade-offs. Transparent transducer designs, such as those utilizing polyvinylidene fluoride (PVDF) films with silver nanowire electrodes or PVDF film coated with indium tin oxide (ITO), enable coaxial optical and acoustic paths but inevitably attenuate incident laser pulse energy (up to $\sim$40$\%$ at 532\si{\nano \meter}) \cite{zhu2024high} \cite{fang2019focused} \cite{jiaming2025high}. 

Alternative configuration using parabolic mirror to achieve acoustic focus, however, this approach introduces alignment complexity and signal loss \cite{zhang2012reflection}. Some systems use prism-based design to redirect acoustic signal, creating comparatively large setup less suitable for compact integration \cite{maslov2008optical}. Some systems utilize miniature transducers (<1\si{\milli \meter}) directly attached to low NA refractive objectives, achieving short working distances with limited solid angle of acoustic detection \cite{liu2019integrated}. In another approach, the laser beam is focused through the central aperture of a ring-shaped transducer using customized water-immersion objectives \cite{cao2023label}. Although this configuration preserves optical performance, the annular transducer geometry inherently creates an acoustic blind spot around the central axis, thereby limiting acoustic sensitivity along the optical axis.

Beyond its high spatial resolution, OR-PAM offers important advantages in terms of system integration. Its compact and flexible design facilitates straightforward combination with other optical microscopy modalities, such as fluorescence lifetime imaging (FLIM), enabling multimodal imaging that integrates molecular contrast with functional information. In such multimodal configurations, it is particularly important that the performance of the optical microscopy modality is not compromised.

In this work, we present an optimized reflection-mode OR-PAM system that addresses these limitations through an integrated configuration. By using a reflective objective, the spatial constraints of conventional systems are reduced, enabling the integration of a large area PVDF transducer within the optical obscuration zone. The transducer is centrally mounted on a meniscus lens, providing a stable water coupling layer that preserves the optical path while ensuring efficient acoustic transmission. The water coupled interface between the transducer and the sample follows established practices in acoustic microscopy to achieve effective acoustic impedance matching. We further present a comprehensive characterization of the experimental setup, including spatial resolution analysis, determination of the laser pulse energy at the sample plane, evaluation of the photoacoustic signal as a function of laser pulse energy, and comparison of the acquired photoacoustic images with corresponding optical microscopy and H\&E stained histological images.

\section{Methods}

\begin{figure}[htbp]
\centering
\includegraphics[width=1\textwidth]{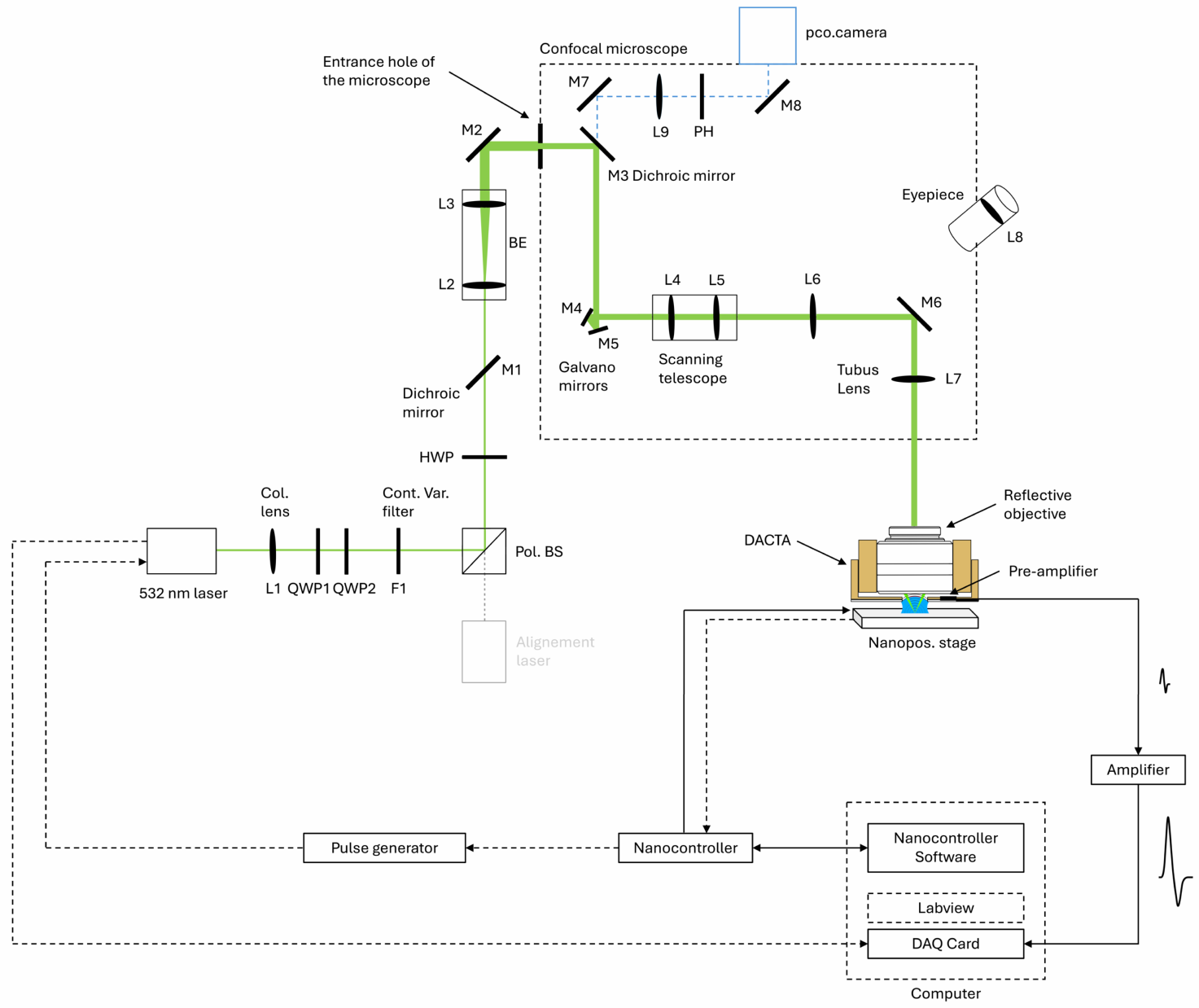}
\caption{Schematic of the OR-PAM experimental setup based on a reflective objective, which includes the hardware part of the optical path, the signal amplification and the data acquisition part. Here L are the lenses, QWP are the quarter-wave plates, F are the filters, HWP is the half-wave plate, M are the mirrors, BS is the beam splitter, BE is the beam expander and, the Distortion-Free Acoustically Coupled Transducer (DACTA).
}
\label{fig:Optical_Path_Laser}
\end{figure}

Figure \ref{fig:Optical_Path_Laser} shows the schematic of the reflection-mode Optical-Resolution Photoacoustic Microscopy (OR-PAM) setup, including the optical path, signal amplification, and data acquisition components.

\begin{figure}[htbp]
\centering
\includegraphics[width=0.6\textwidth]{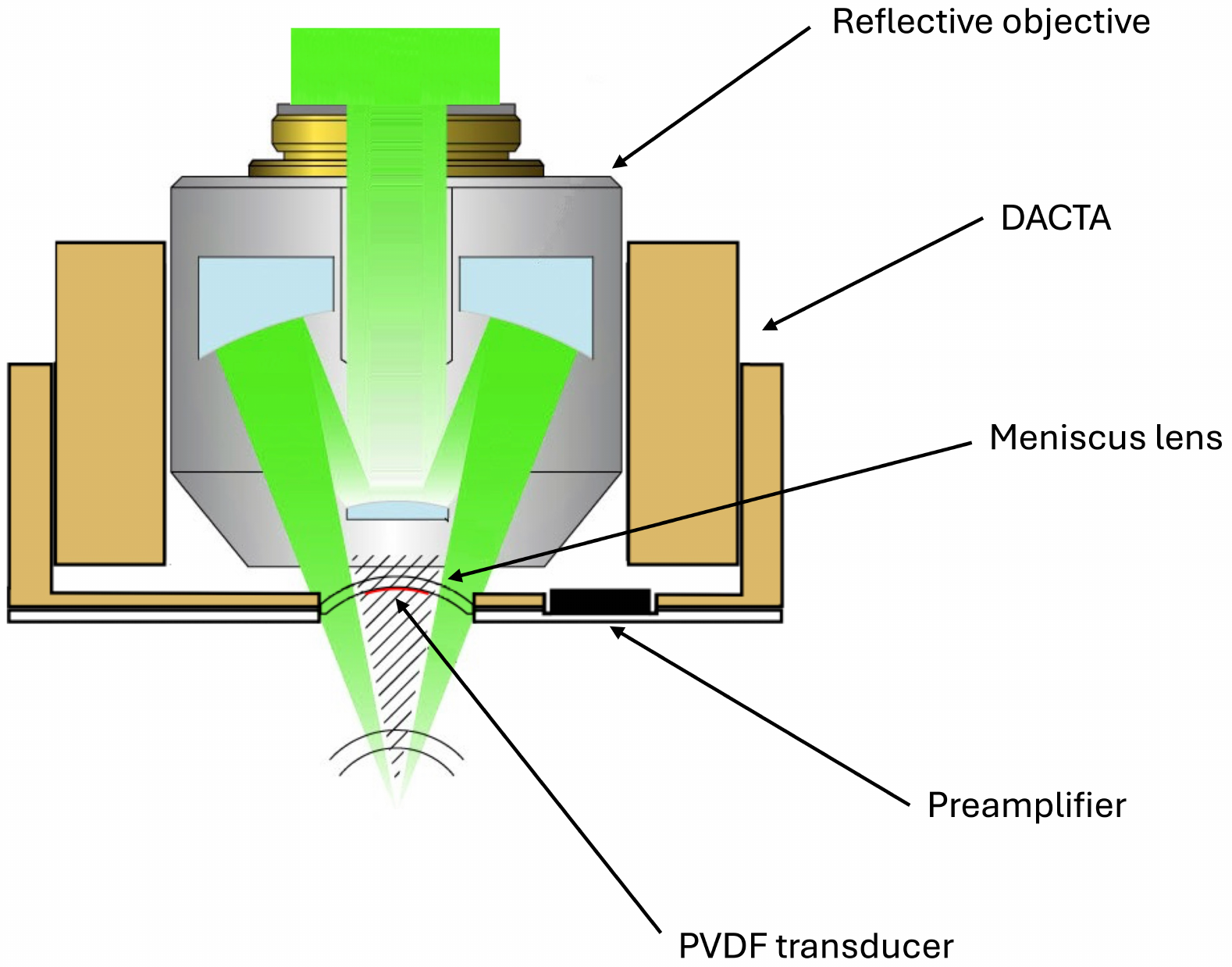}
\caption{Close-up view of the DACTA (Distortion-Free Acoustically Coupled Transducer), with representation of its constituents parts.}
\label{fig:DACTA}
\end{figure}

The laser source is the Cobolt Tor XE ($532\si{nm}$ wavelength, pulse energy $250 \pm 25$\si{\micro \joule}, pulse width $2 \pm 1\si{\nano \second}$, repetition rate $1\si{\kilo \hertz}$), a triggerable, high performance Q-switched diode-pumped laser, chosen for its high pulse-to-pulse stability \cite{Cobolt_manual}. The emitted beam is first collimated by lens L1 to a diameter of $\sim 1.5\si{\milli \meter}$. Two quarter-wave plates (QWP1, QWP2) are used to adjust the polarization state (S-polarization) to optimize reflection at the polarizing beam splitter (pol. BS), thereby directing the light into the microscope and maximizing the optical power coupled into the desired beam path. A continuously variable ND filter (F1, NDC-50C-4M, Thorlabs; OD range: 0.04–4.0) has been placed to control the input laser beam flux. The beam is then expanded using a beam expander (BE) to match the entrance pupil. The output of the beam expander is a wider quasi-Gaussian beam. The expanded Gaussian beam is reflected by mirror M2 into the microscope structure, after passing through a $\sim7\si{\milli \meter}$ circular opening, then it is reflected by mirrors M3–M6, and is focused on a small spot on the sample by a reflective objective (LMM-40X-UVV, Thorlabs; entrance pupil: $5.1\si{\milli \meter}$; working distance: $7.6\si{\milli \meter}$; magnification: $40\times$; nominal NA: $0.5$). This focused beam produces the acoustic signal through the photoacoustic effect. The central obscuration of the reflective objective allows the integration of acoustic receiver within the optical pathway without introducing additional optical distortion, a core innovation of our study that allows confocal alignment of optical and acoustic path with large angular aperture in a compact Distortion-Free Acoustically Coupled Transducer (DACTA), as shown in Fig.\ref{fig:DACTA}.

The DACTA assembly comprises two brass parts: a base fixed to the objective and an outer housing that screws onto it. The housing has a central hole sealed by a custom meniscus lens (thickness: $0.7\si{\milli \meter}$), supporting a centered PVDF transducer (polyvinylidene fluoride, $5\si{\milli \meter}$ diameter, $20\si{\micro \meter}$ thickness; manufactured by the Pavel Subchev group, Nizhny Novgorod, Russia). This design utilizes the obscuration zone of the reflective objective to position the transducer without obstructing the laser path. A built-in preamplifier amplifies the weak PA signals before it is transported by cable to the second amplifier. This step minimizes noise introduced by the cable. In this study, raster scanning is implemented by mounting the sample on a piezoelectric translation stage (Nano-LPS200, Mad City Labs; nanopositining stage) that provides precise 2D raster scanning (controlled by the 3-axis control system \textit{MCL nano-drive 3} (ND3-USB203-AR5-ISS, Mad City Labs). However, mechanical scanning is not an inherent limitation of the proposed setup and can be replaced by galvo mirrors to achieve faster scanning speeds. All components are synchronized via dedicated control software, wherein the positioning signals from the nanopositioning stage are used to trigger the pulse generator (TG5011, Aim-TTi). The pulse generator then delivers a predefined number of trigger pulses to the laser system to generate laser pulses for each pixel throughout the scan. 

The scanning area is 200 $\times$ 200 \si{\micro \meter \squared}, with 400 measurement points on each axis and 128 pulse signals averaged per pixel to enhance SNR. The PA signals are amplified via a low-noise, broadband, RF amplifier (AU-1442-400, Miteq) and sampled with $250\si{\mega \hertz}$. In this setup, water is used as acoustic coupling medium between sample and PVDF transducer. 

The scan data are acquired and stored using the control software operating the system and are subsequently processed using a customized MATLAB code developed specifically for this OR-PAM setup. For both pre-scan sample alignment and post-scan comparative imaging, the system uses a scientific CCD camera (PCO.1600, $1600 \times 1200$ pixels) in combination with a refractive objective (LD Achroplan $40\times$ $/0.60$, ZEISS). This optical subsystem provides complementary visualization to the OR-PAM based acoustic imaging.

In this study, sections from B16F10 tumors implanted in mice were provided by the Institute of Tissue Medicine and Pathology of the University of Bern, and prepared using standard histopathology protocols: paraffin-embedded, microtome-sectioned, and mounted on microscope glass slides after warm water flattening.

\section{Results}

\subsection{Evaluation and characterization of the imaging resolution}
\label{sub:characterisation_imaging_resolution}

A critical performance metric of the OR-PAM system is its spatial resolution, which is quantified using a USAF 1951 resolution target. The characterization comprises two components: axial and lateral resolution.

\begin{figure}[htbp]
\centering
\begin{subfigure}{0.4\textwidth}
    \centering
    \includegraphics[width=\textwidth]{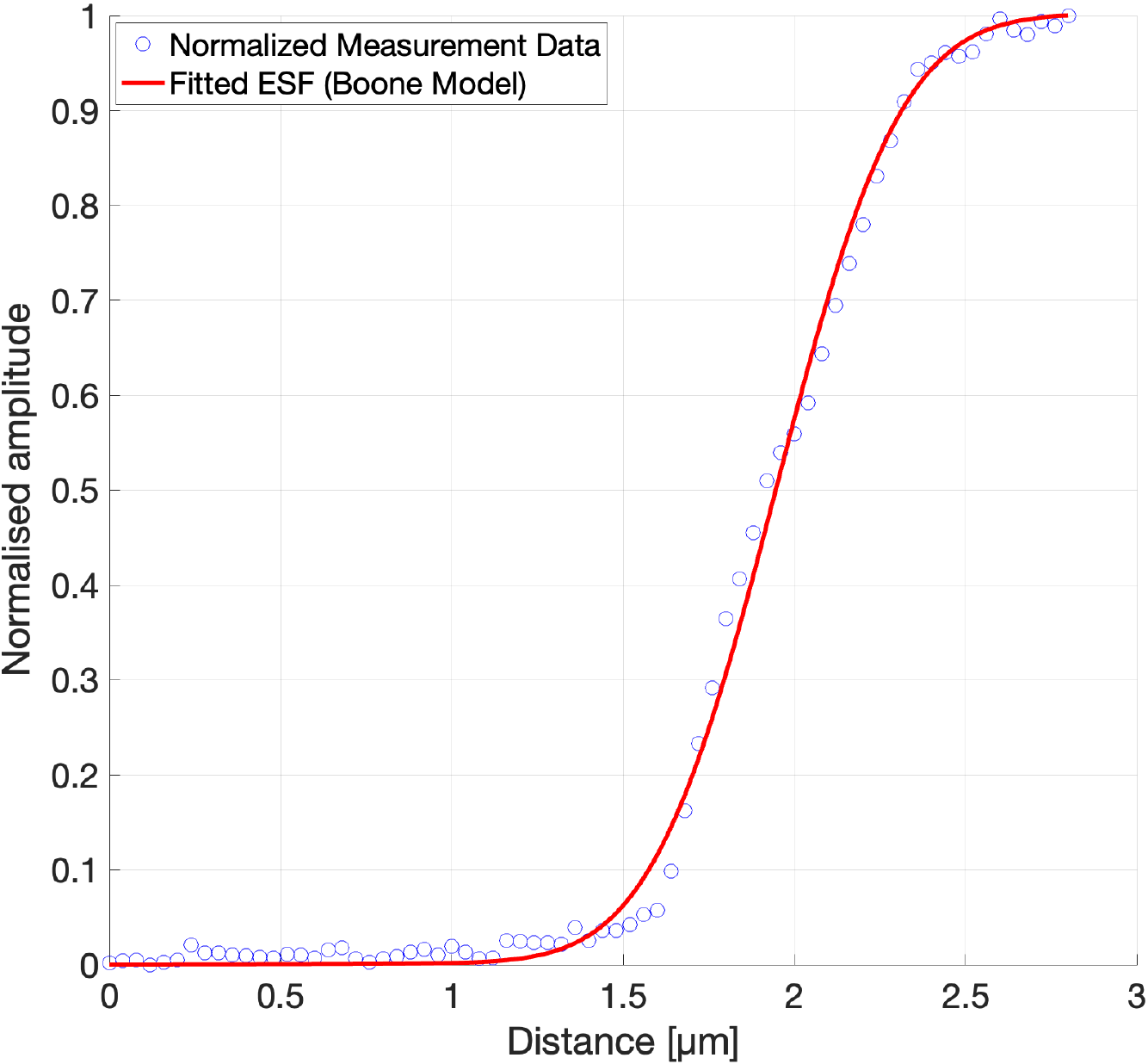}
    \caption{}
    \label{fig:esf}
\end{subfigure}
\hfill
\begin{subfigure}{0.4\textwidth}
    \centering
    \includegraphics[width=\textwidth]{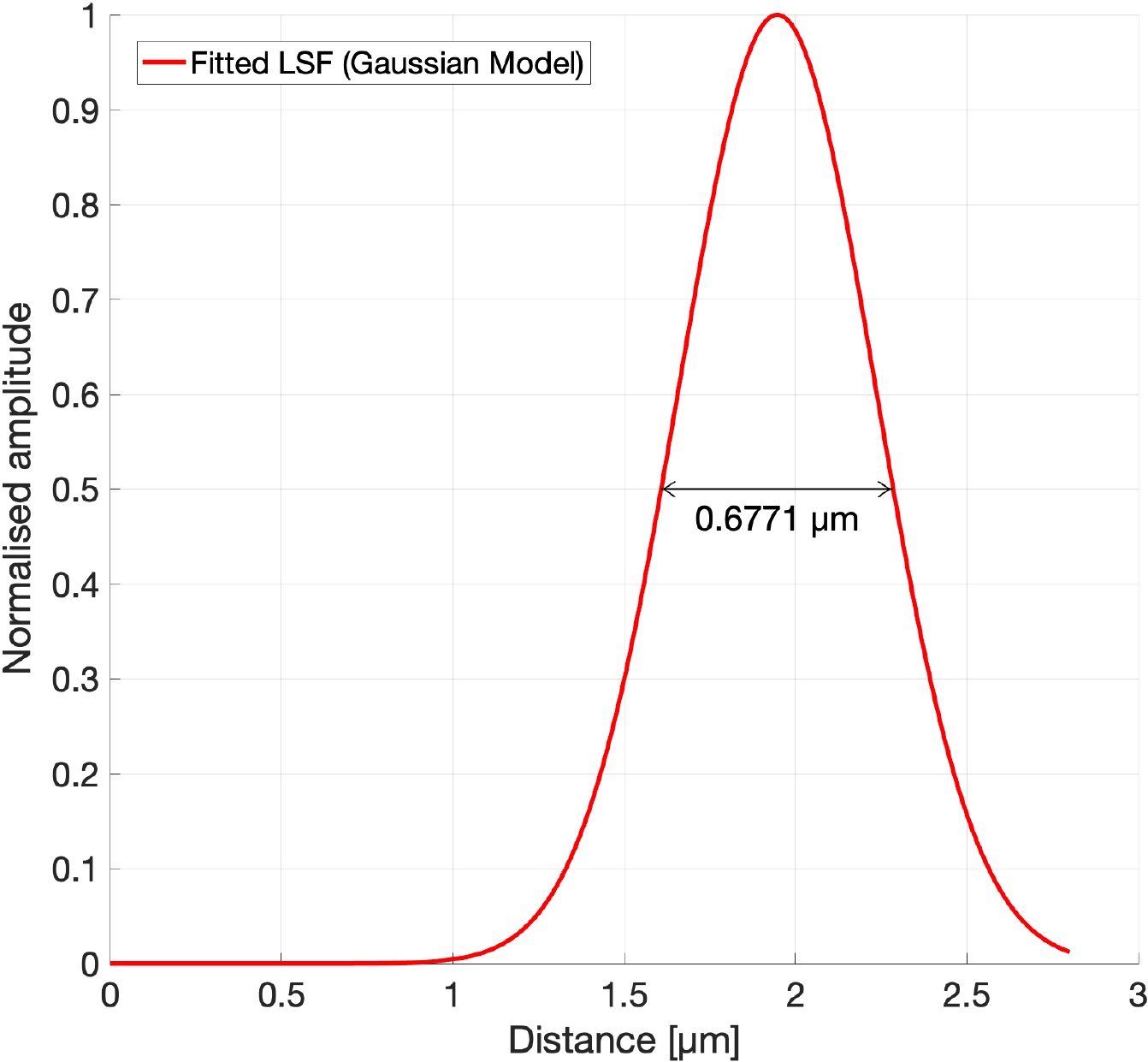}
    \caption{}
    \label{fig:lsf}
\end{subfigure}
\caption{Lateral resolution characterization of the OR-PAM system: (a) Edge spread function (ESF) measured at a sharp edge of a USAF 1951 resolution target (blue dots: experimental data averaged over 100 scans; red line: theoretical fit following Boone et al. model \cite{John_M_Boone}). (b) Line Spread Function (LSF) derived from the ESF with Gaussian fit, showing the FWHM value used for resolution quantification.}
\label{fig:esf_boone}
\end{figure}

 Lateral resolution was evaluated over a sharp edge of the USAF 1951 target.
The stage was fixed along y and z, while x was positioned 1.5\si{\micro \meter} before the chrome-painted bar edge. The edge spread function (ESF) was derived by 100 scans (0.04\si{\micro \meter} steps over 3\si{\micro \meter}) across the edge. The averaged data were normalized and fitted using the algorithm presented by Boone and Seiber \cite{John_M_Boone}, as shown in Fig.\ref{fig:esf}, with discrete measurements (blue dots) and model fit (red line). The line spread function (LSF) was derived by the ESF, with its FWHM defining the lateral resolution of the system Fig.\ref{fig:lsf}. 

 The diffraction-limited optical lateral resolution can only be estimated approximately in the current setup, as it is influenced by  experimental factors. Therefore, the lateral resolution of the OR-PAM system is determined by the diffraction-limited spot size of the optical focus and can be expressed by the equation \cite{https://doi.org/10.1002/lpor.201200060}\cite{NA}:

\begin{equation}
R_{L,OR} = 0.51 \frac{\lambda}{NA}.
\label{eq:R_{L,OR}}
\end{equation}

 For an ideal Gaussian beam and a numerical aperture (NA) of $0.5$ for the focusing lens, the optical lateral resolution calculated by Eq. \ref{eq:R_{L,OR}} is $0.54\si{\micro \meter}$. The experimentally determined lateral resolution is $0.67 \si{\micro\meter}$.

\subsection{Dependence of PA signal on pulse energy}
\label{sub:laser_beam_intensity_analysis}

An important aspect we analyzed was the consistency of the signal shape with respect to pulse energy, which allowed us to establish a relationship between laser pulse energy, signal amplitude, and the temporal profile of the measured photoacoustic (PA) signal. To investigate this, we selected a specific point on a biological sample and performed repeated measurements at the same location.

The acquired PA signals were processed using a MATLAB code. The signals were band-pass filtered in the $2.1 - 45$ MHz rage using Fourier-domain filtering method to isolate the PA signal while suppressing out-of-band noise contribution. This signal-processing approach effectively enhanced the signal-to-noise ratio (SNR), particularly for low-amplitude signals, while preserving the essential temporal characteristics of the PA response.

\begin{figure}[htbp]
\centering
\begin{subfigure}{\textwidth}
    \centering
    \includegraphics[width=0.7\textwidth]{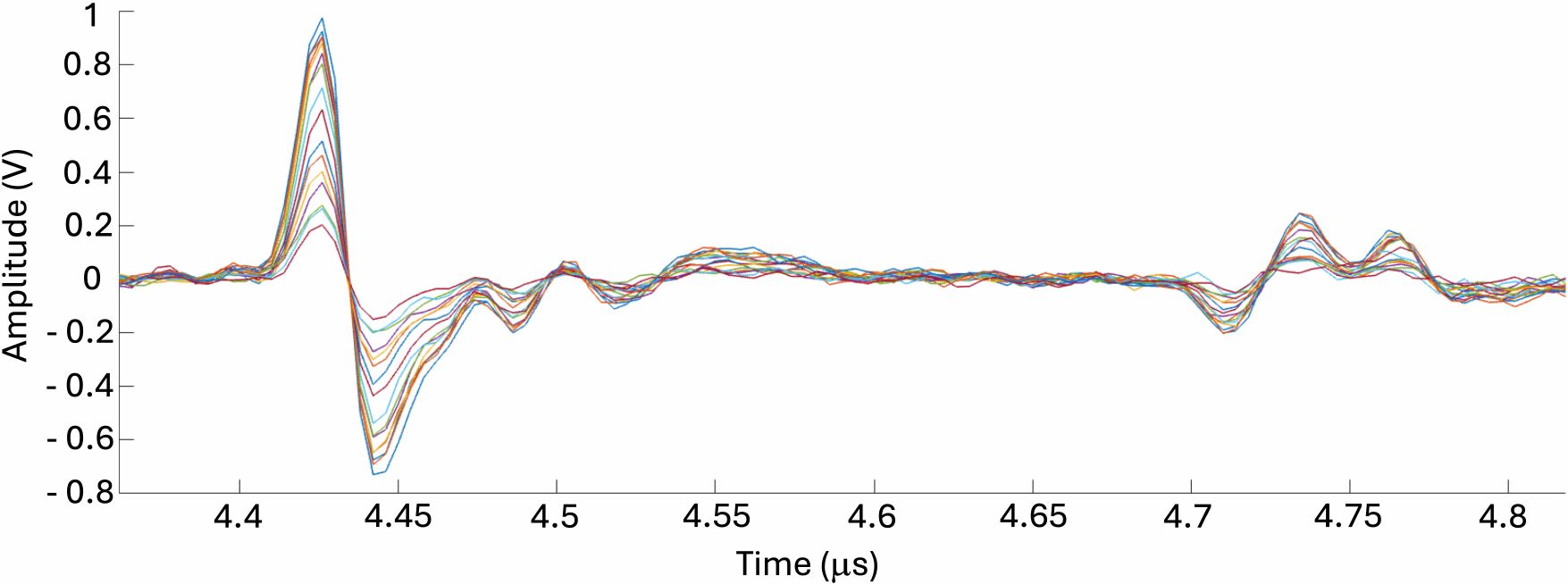}
    \caption{}
    \label{fig:single_point_variation_all}
\end{subfigure}
\vspace{1em} % Adjust space between figures if necessary
\begin{subfigure}{\textwidth}
    \centering
    \includegraphics[width=0.7\textwidth]{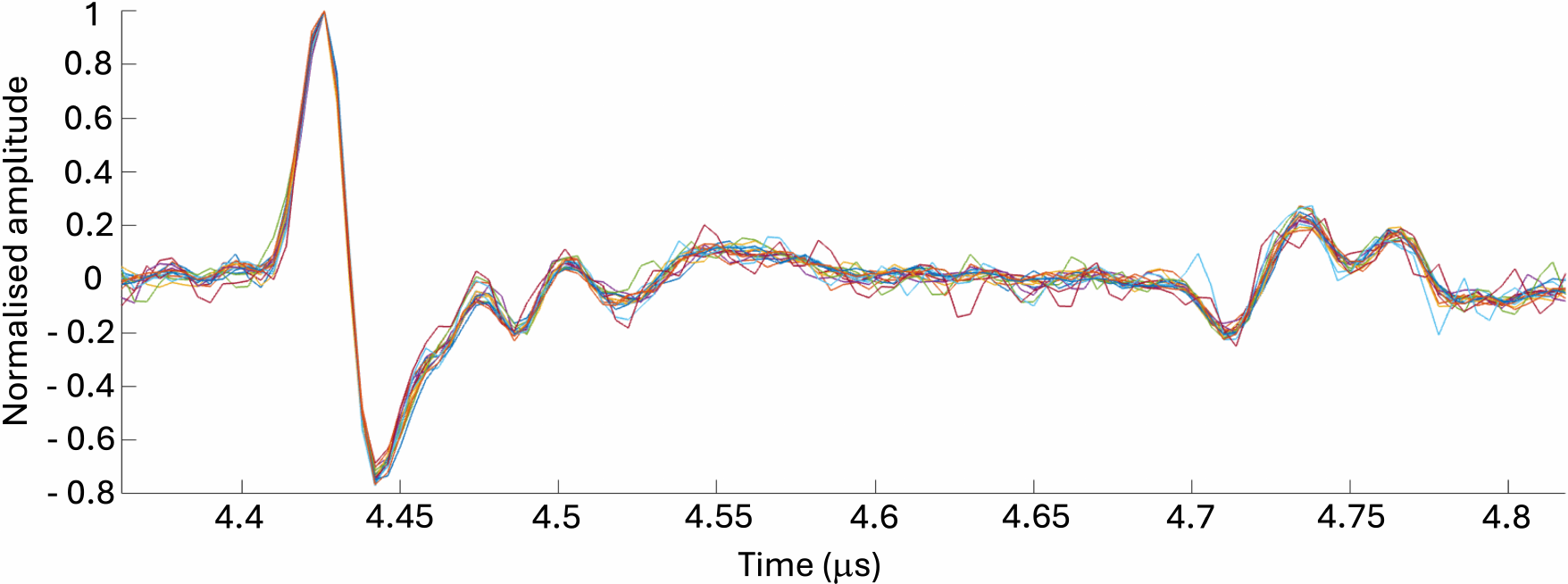}
    \caption{}
    \label{fig:single_point_variation_normalised}
\end{subfigure}
\caption{Photoacoustic signal response to laser pulse energy variation ($1nJ$ to $100nJ$). (a) Signal amplitudes $[\si{\volt}]$ decrease with reduced pulse energy. (b) Normalized signals show identical temporal waveform structure across decreasing pulse energy, confirming pulse energy-invariant thermoelastic properties.}
\label{fig:single_point_variation}
\end{figure}

At a laser pulse energy of $100 nJ$, the PA signal approached the saturation limit of the detection system  $0.5\si{\volt}$, see Fig.\ref{fig:single_point_variation_all}, indicating optical absorption at the measurement site. As the laser pulse energy was reduced, the signal amplitude decreased linearly while preserving the overall waveform morphology. This behavior was confirmed through comparison of the normalized signals shown in Fig.\ref{fig:single_point_variation_normalised}, demonstrating that the thermoelastic response remained invariant over the investigated pulse energy range. The preservation of the waveform shape across varying pulse energies indicates that high-fidelity photoacoustic imaging can be achieved at the pulse energies used for this study. This capability is particularly important for minimizing optical exposure and reducing the risk of photothermal or photomechanical damage to biological samples.

\subsection{Correlating Photoacoustic and Histological features in sections from B16F10 tumors implanted in mice}
\label{sub:Photoacoustic_Histological}

\begin{figure}[htbp]
\centering
\begin{subfigure}{0.32\textwidth}
    \centering
    \includegraphics[width=\textwidth]{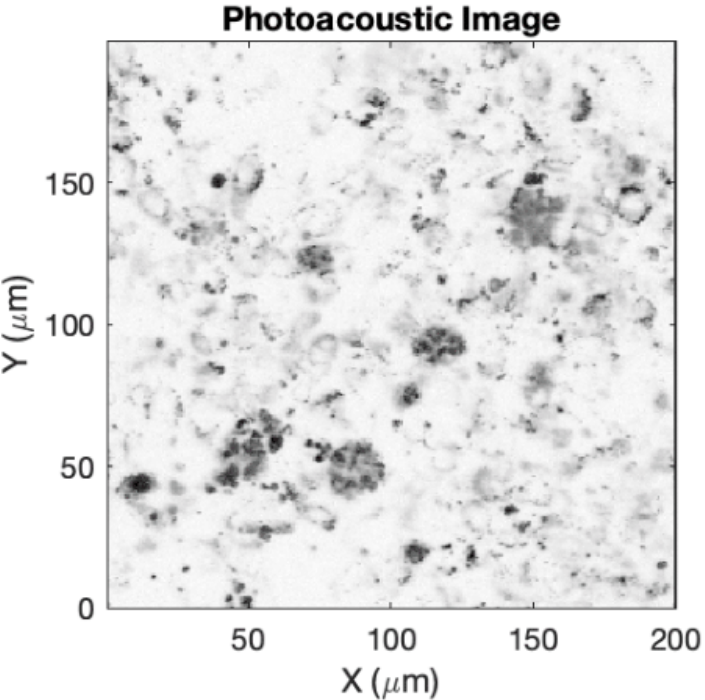}
    \caption{}
    \label{fig:pa}
\end{subfigure}
\hfill
\begin{subfigure}{0.32\textwidth}
    \centering
    \includegraphics[width=\textwidth]{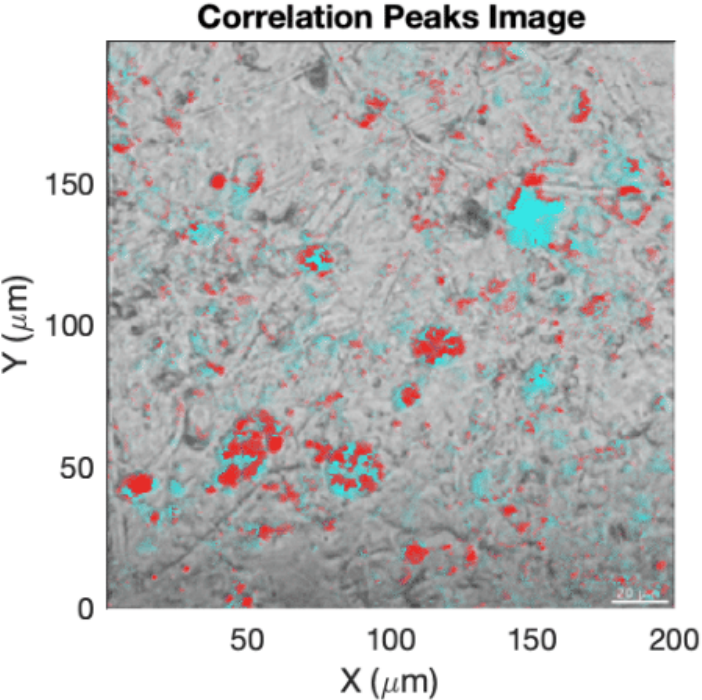}
    \caption{}
    \label{fig:correlation}
\end{subfigure}
\hfill
\begin{subfigure}{0.32\textwidth}
    \centering
    \includegraphics[width=\textwidth]{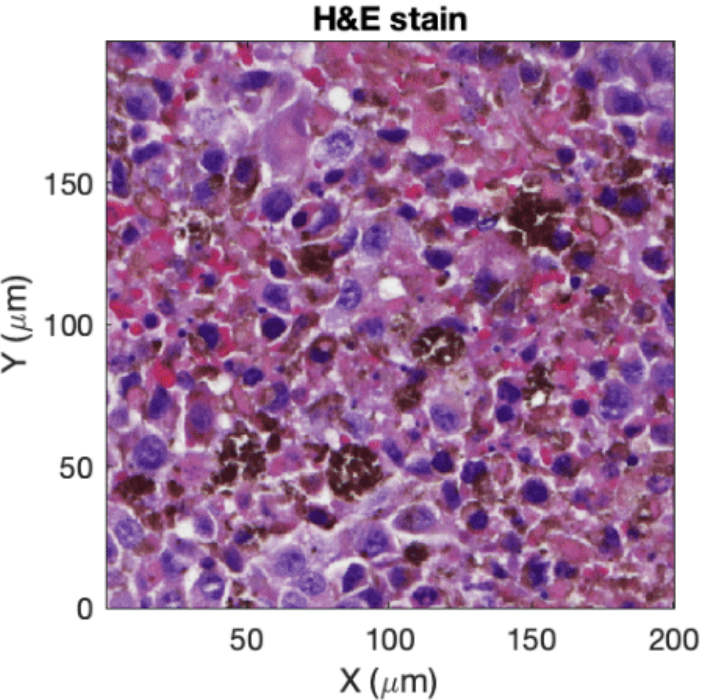}
    \caption{}
    \label{fig:he}
\end{subfigure}
\caption{Multi-modal image display of sections from B16F10 tumors implanted in mice : (a) Photoacoustic image, (b) Optical microscopy image overlaid with peak ratio classes ($\Delta < 1.35$: cyan; $\Delta \geq 1.35$: red), and (c) $H\&E$ stained histology image. All panels show the same $200 \times 200 \si{\micro \meter \squared}$ sample region.}
\label{fig:three_13}
\end{figure}

Photoacoustic images were reconstructed by assigning the peak amplitudes of the band-pass filtered signals to their corresponding scan positions within a spatial matrix. The resulting images are displayed using an inverted grayscale representation, in which the maximum signal amplitude is encoded in black and zero amplitude in white Fig.\ref{fig:pa}. The reconstructed OR-PAM images, covering a field of view of $200 \times 200 {\mu m}^2$, were spatially aligned with optical microscopy images $233 \times 311 {\mu m}^2$ using an automated feature-matching algorithm implemented in the processing code. In cases where automatic alignment was unsuccessful, such as for low-contrast features, manual landmark selection was performed. This coregistration procedure enabled direct spatial comparison between the optical microscopy images and photoacoustic images within the same sample region.

 To systematically evaluate $532 nm$ laser interactions with biological samples, we performed OR-PAM scans over regions exhibiting a uniform cellular distribution in field-of-view of $200 \times 200 {\mu m}^2$ and $300 \times 300$ pixels with $0.67 \mu m$ step size. The pulse energy was fixed at $60 nJ$ to preserve sample integrity. For histological correlation, the samples were stained with H\&E which is done by the Institute of Tissue Medicine and Pathology under Prof. Dr. phil. nat. Mirjam Schenk, and imaged using a Plan-Apochromat $40\times$ objective on a Pannoramic Slide Scanner (pixel size: $0.1215 \mu m/pixel$) Fig.\ref{fig:he}. Comparative analysis between the H\&E stained sections and the photoacoustic images showed consistent spatial agreement between regions of strong photoacoustic signal and histologically identified melanin distribution. Although H\&E staining is not specific to melanin, the characteristic black-brown pigmentation of these features, as confirmed through consultation with clinical and biomedical experts, strongly suggests melanin as the dominant absorber at $532 nm$. The demonstrated sensitivity to melanin highlights the capability of OR-PAM for label-free mapping of melanin distribution.

\begin{figure}[htbp]
\centering
\includegraphics[width=0.5\textwidth]{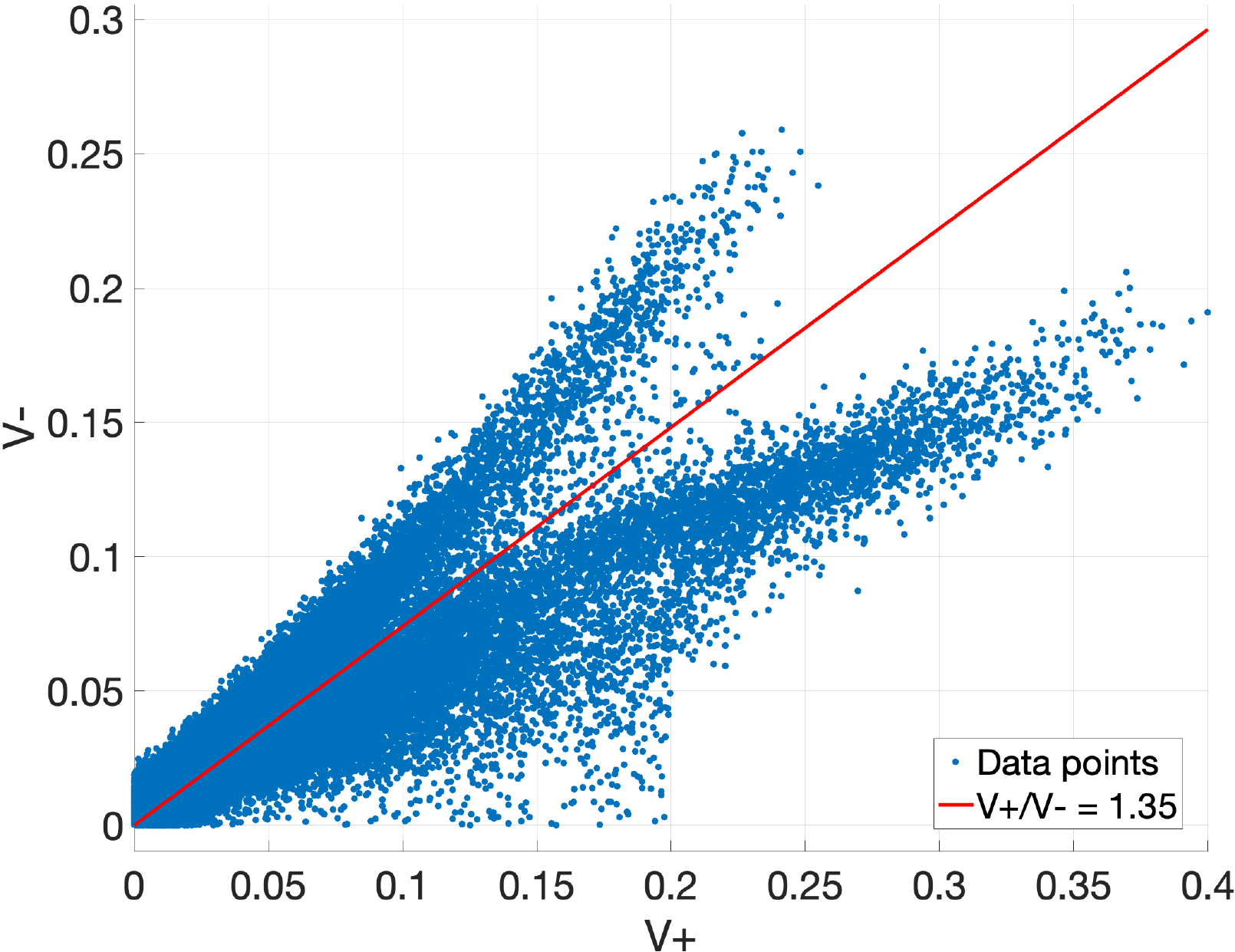}
\caption{Relation between positive '$V+$' and negative '$V-$' photoacoustic peak amplitudes across all image pixels. The red line indicates the classification threshold at $\Delta = 1.35$.}
\label{fig:V+_V-}
\end{figure}

When analyzing the photoacoustic signals, we observed a bipolar waveform shape that remained independent of the laser pulse energy but varied across different scan positions. To characterize this behavior across scans, the negative '$V-$' and positive '$V+$' peak amplitudes for all pixels were plotted against each other Fig. \ref{fig:V+_V-}. This systematic analysis revealed two distinct classes of peak behavior. To enable classification, we defined a classification parameter $\Delta$:

\begin{equation}
    \Delta = \bigg| \frac{\text{Positive peak amplitude}}{\text{Negative peak amplitude}}\bigg| = \bigg| \frac{V^+}{V^-} \bigg|.
\end{equation}

 A value of $\Delta$ = 1.35 separates the two peak classes. 

 Using this criterion, two filtered matrices were generated: one containing pixels with $\Delta < 1.35$, where the corresponding values were retained while all other entries were set to NaN, and a second matrix containing pixels with $\Delta \geq 1.35$, with all remaining values similarly set to NaN. The assignment of NaN values ensured transparency when these matrices were overlaid onto optical or photoacoustic images. To visualize the spatial distribution of the two peak classes, a custom color map was applied: transparent-to-cyan mapping for $\Delta < 1.35$ and transparent-to-red mapping for $\Delta \geq 1.35$. As ahown in Fig. \ref{fig:correlation}. This overlay on the optical microscopy image reveals distinct spatial patterns in the tissue distribution of the two peak classes.

 Peak amplitudes analysis demonstrated distinct maxima between classes: $0.255 \si{\volt}$ for $\Delta < 1.35$ spectra versus $0.400 \si{\volt}$ for $\Delta \geq 1.35$ spectra, suggesting potential differences in absorber properties or local thermoelastic environments. Across repeated measurements of the same mice sample, the $\Delta$-based classification exhibited high consistency, confirming a non-random distribution of the two peak classes. However, correlation with H\&E stained melanin patterns did not resolve this dichotomy, as the staining lacked specificity to differentiate them. Future studies by using melanin-specific histological methods could clarify the origin of these distinct peak classes.

\section{Conclusion}

The proposed approach provides improved signal detection while maintaining the compact integration requirements essential for optical microscopy applications. The presented configuration combines three principal advantages: preservation of optical performance through a reflective objective-based design, improved acoustic sensitivity achieved by integrating a large-area transducer, and practical compatibility with standard microscopy platforms. Although the present implementation is limited to a single excitation wavelength, the system can be readily extended to multi-wavelength operation, which would enable improved discrimination between different classes of optical absorbers. We used mechanical scanning for this work but the system can also be extended to optical scanning using galvanometer mirrors to enable higher frame-rate imaging. Together, these features significantly advance the capabilities of OR-PAM for high-resolution biomedical imaging applications.

\section*{Acknowledgment}

The author acknowledges the continuous guidance and feedback of Dr. Michael Jaeger throughout the research and during the review of the results. The author also thanks the Institute of Tissue Medicine and Pathology at the University of Bern for providing the biological samples and for their support with histological preparation. Further appreciation is extended to colleagues and collaborators for valuable discussions and technical assistance during the development and characterization of the OR-PAM system.

%%%%%%%%%%%%%%%%%%%%%%% References %%%%%%%%%%%%%%%%%%%%%%%%%
\bibliography{sample}

\end{document}